\documentclass[aps,prx,twocolumn,superscriptaddress,showpacs,floatfix]{revtex4-2}
\usepackage{mathrsfs}
\usepackage{amsmath,amssymb,graphicx}
\usepackage{wasysym}
\usepackage[utf8]{inputenc}
\usepackage[T1]{fontenc}
\usepackage{xcolor}
\allowdisplaybreaks
\usepackage{rotating} %% for dimer
\usepackage[columnwise]{lineno}
\usepackage{verbatim}

\usepackage{textcomp}

\usepackage{bm}

\IfFileExists{siunitx.sty}{\usepackage{booktabs,siunitx}}{}

\usepackage{color}
\definecolor{LinkColor}{rgb}{0.256,0.439,0.588}
\usepackage{hyperref}
\hypersetup{
    pdfstartview={FitH},    % fits the width of the page to the window
    colorlinks=true,       % false: boxed links; true: colored links
    linkcolor=blue,          % color of internal links (change box color with linkbordercolor)
    citecolor=blue,        % color of links to bibliography
    filecolor=magenta,      % color of file links
    urlcolor=blue           % color of external links
}

\renewcommand{\vec}[1]{\bm{#1}}
\newcommand{\bra}[1]{\langle#1\rvert}
\newcommand{\ket}[1]{\lvert#1\rangle}

\usepackage{pifont}

\newcommand{\La}{\line (1,0  ){5}}      % first plaquette  bond1 location longth
\newcommand{\Ld}{\line (-2,0){5}}     % first plaquette  bond2 location longth
\newcommand{\C} {\circle{3}}        % diameter of circle

\newcommand{\LaT}{\rule[-0.7pt]{0.18cm}{0.15em}}  %% Bold   % second plaquette  bond1 location longth
\newcommand{\LdT}{\rule[-0.7pt]{0.18cm}{0.15em}}  %% Bold    second plaquette  bond2 longth width
\newcommand{\LbT}{\rotatebox{60}{\rule[-1pt]{0.18cm}{0.15em}}}  %% Bold   first plaquette  bond4 longth width
\newcommand{\LeT}{\rotatebox{60}{\rule[-1pt]{0.18cm}{0.15em}}}  %% Bold    first plaquette  bond3 longth width

\newcommand{\LbTs}{\rotatebox{60}{\rule[1.2pt]{0.18cm}{0.05em}}}  %%  second plaquette  bond4 longth width
\newcommand{\LeTs}{\rotatebox{60}{\rule[2.3pt]{0.18cm}{0.05em}}}  %%  second plaquette  bond3 longth width

\newcommand{\pZ}{\put(-0.5,0)}    % left upper site location
\newcommand{\pC}{\put(7.5,0)}   % right upper site location
\newcommand{\pA}{\put(-4,-7)}    % left lower site location
\newcommand{\pB}{\put(4,-7)} % right lower site location

\newcommand{\pZb}{\put(0.8,0)}    % left upper bond location
\newcommand{\pCb}{\put(6,0)}   % right upper bond location
\newcommand{\pAb}{\put(-2.5,-7)}    % left lower bond location
\newcommand{\pBb}{\put(4.3,-6.5)} % right lower bond location

\newcommand{\pAT}{\put(-4.3,-5.5)} %%for thickness dimer      first plaquette  bond3 location longth
\newcommand{\pBT}{\put(3.7,-5.5)}  %%for thickness dimer    first plaquette  bond4 location longth

\newcommand{\rhomb}{            % first four sites
  \pA{\C}\pB{\C}\pZ{\C}\pC{\C}
 }

\newcommand{\rhombH}{            %secondt plaquette
  \begin{picture}(20,5)(-8,-6)
    \pAb{\LaT}\pBb{\LbTs}\pA{\LeTs}\pZb{\LdT}%\pZ{\Lf}
    \rhomb
  \end{picture}
}
\newcommand{\rhombV}{            %first plaquette
  \begin{picture}(20,5)(-8,-6)      % location
   \pAb{\La}\pBT{\LbT}\pAT{\LeT}\pCb{\Ld}%\pZ{\Lf}
    \rhomb
  \end{picture}
}

\graphicspath{./figures}

\def\be{\begin{equation}}
\def\ee{\end{equation}}
\def\bea{\begin{eqnarray}}
\def\eea{\end{eqnarray}}
\def\Tr{{\rm Tr}}

\begin{document}
\title{Symmetrizing the Constraints -- Density Matrix Renormalization Group for Constrained Lattice Models}
% application to triangular lattice quantum dimer model}
\author{Ting-Tung Wang}
\affiliation{Department of Physics and HK Institute of Quantum Science \& Technology, The University of Hong Kong, Pokfulam Road, Hong Kong}

\author{Xiaoxue Ran}
\affiliation{Department of Physics and HK Institute of Quantum Science \& Technology, The University of Hong Kong, Pokfulam Road, Hong Kong}

\author{Zi Yang Meng}
\email{zymeng@hku.hk}
\affiliation{Department of Physics and HK Institute of Quantum Science \& Technology, The University of Hong Kong, Pokfulam Road, Hong Kong}

\date{\today}
%%%%%%%%%%%%
\begin{abstract}
We develop a density matrix renormalization group (DMRG) algorithm for constrained quantum lattice models that successfully {\it{implements the local constraints as symmetries in the contraction of the matrix product states and matrix product operators}}. Such an implementation allows us to investigate a quantum dimer model in DMRG for any lattice geometry wrapped around a cylinder with substantial circumference. We have thence computed the ground state phase diagram of the quantum dimer model on triangular lattice, with the symmetry-breaking characteristics of the columnar solid phase and $\sqrt{12}\times\sqrt{12}$ valence bond solid phase fully captured, as well as the topological entanglement entropy of the $\mathbb{Z}_2$ quantum spin liquid phase that extends to the RK point on non-bipartite lattice accurately revealed. Our DMRG algorithm on constrained quantum lattice models opens new opportunities for matrix and tensor-based algorithms for these systems that have immediate relevance towards the frustrated quantum magnets and synthetic quantum simulators.
\end{abstract}
%%%%%%%%%%%%
\maketitle
%%%%%%%%%%%%
\section{Introduction}
%%%%%%%%%%%%
Frustration arises in systems where all the interactions cannot be simultaneously minimized due to competing interactions, geometric constraints, or inherent randomness and disorder, which often leads to local constraints that the system must satisfy. For example, artificial spin ice~\cite{baxter2007exactly,castelnovoMagnetic2008,gingras2010spin,Nisoli2013Artificial,Bramwell_2020}, which approximates classical frustrated Ising magnets, obey an ice rule that the number of spins pointing inward of a vertex must equal the number of spins pointing outward. These systems facilitate experiments on novel physical phenomena, including vertex-based frustration~\cite{gilbert2014emergent} and the emergence of magnetic monopole excitations~\cite{castelnovoMagnetic2008,ladak2010direct}. Similarly, in fully packed quantum dimer or loop models~\cite{moessner2010quantum}, which share low energy properties with the frustrated quantum Ising~\cite{moessnerIsing2001,moessnerShort2001,moessnerResonating2001,wangCaution2017}, XXZ~\cite{balentsFractionalization2002,wangQuantum2018,sunDynamical2018,wangFractionalized2021} or Bose-Hubbard-type~\cite{isakvoHardcore2006,isakovTopological2011,Isakov2012} models, only one or two dimers can occupy each lattice site. These systems often exhibit exotic phenomena such as fractionalization, topological order, unconventional phase transitions, and long-range quantum entanglement. However, simulating these systems poses considerable challenges due to intrinsic and computational difficulties. These include the requirement for discrete degrees of freedom to adhere to specific (often strictly local) constraints and the presence of highly degenerate ground states resulting from frustration. These factors collectively complicate the exploration of their phase space and the accurate characterization of their emergent properties.

As prototypical constrained many-body systems, quantum dimer/loop models (QDM/QLM) are characterized by the local constraint with dimer coverings, where one dimer (for QDM) and two dimers (for QLM) occupy each site. These models have played a pivotal role in advancing our understanding of quantum spin liquids~\cite{moessnerIsing2001,Moessner2001b, moessnerShort2001,moessnerResonating2001,Ralko2005,Ralko2006,Ralko2007,YanTopological2021,ZYan2022,Rousochatzakis2014Quantum,Vernay2006Identification}  and excitations of fractionalized quasiparticle~\cite{Ivanov2004,Misguich2008,YanTopological2021,ZYan2022}, effectively describing the complex phenomena observed in frustrated magnets~\cite{moessnerIsing2001, Moessner2001b, moessnerShort2001, Jiang2005String} and cold atom, Rydberg array and quantum simulator experiments~\cite{Glaetzle2014Quantum,Bernien17,Samajdar:2020hsw,Verresen:2020dmk,Semeghini21,Giudici2022Dynamical}. In QDMs, the $\mathbb{Z}_2$ quantum spin liquid (QSL) phase emerges from the solvable Rokhsar-Kivelson(RK) point~\cite{rokhsarSuperconductivity1988,moessner2010quantum} on nonbipartite triangular~\cite{YanTopological2021,moessnerShort2001,Ralko2007,Ralko2006} and kagome lattice~\cite{Misguich2003Quantum,Rousochatzakis2014Quantum,Hwang2024Vison}. Away from the RK point, phase transitions between $\mathbb{Z}_2$ topological phase and other topologically trivial symmetry breaking phases are frequently observed~\cite{moessnerIsing2001,Moessner2001b, moessnerShort2001,moessnerResonating2001,YanTopological2021,Ralko2005,Ralko2006,Misguich2003Quantum,Rousochatzakis2014Quantum,Hwang2024Vison}. However, due to the challenges associated with simulating in the reduced Hilbert space with local constraints, large-scale numerical studies of these models remain relatively seldom. For QDM on the triangular lattice, the ground-state phase diagram has been explored in a limited number of studies using exact diagonalization (ED)~\cite{Ralko2005} and the quantum Monte Carlo (QMC) method~\cite{moessnerResonating2001,Ivanov2004,Ralko2005,Ralko2006,Ralko2007,YanTopological2021}. An intermediate $\sqrt{12}\times\sqrt{12}$ valence bond state (VBS) was found between the spin liquid and columnar phases. The transition points are determined through extrapolations of the topological gap~\cite{Ralko2005} as well as the excitation gaps in dimer and vison spectra~\cite{Ralko2006,Ralko2007,YanTopological2021}, but the nature of the quantum criticality is still not quantitatively revealed, precisely due to the numerical difficulty mentioned above.

Meanwhile, entanglement entropy (EE) has been widely employed as a powerful tool to identify critical and topological phases in quantum many-body systems~\cite{kitaevTopological2006,levinDetecting2006,isakovTopological2011,Jiang_Identifying_2012,Zhang_Quasiparticle_2012,zhaoScaling2022,zhaoMeasuring2022,liaoExtracting2024,songExtracting2024,songEvolution2025,zhaoUnconventional2025,wangAnalog2025}. In triangular QDM, the $\mathbb{Z}_2$ topological order can be characterized by a non-trivial topological EE, which exhibits a topological term $\gamma=-\ln(2)$ for the cylindrical geometry and $\gamma=-2\ln(2)$ for torus~\cite{levinDetecting2006,kitaevTopological2006,isakovTopological2011,Jiang_Identifying_2012,zhaoMeasuring2022,chenTopological2022,wangAnalog2025}. This conclusion has been proven at the exactly solvable RK point through both theoretical analyses utilizing reduced density matrix~\cite{Stephan_2012Renyi} and numerical studies using ED~\cite{Furukawa2007Topological},  projected
entangled-pair states~\cite{KrishanuTopological2015}, and Monte Carlo methods~\cite{peiRenyi2014}. However, the behavior of EE away from the RK point~\cite{Furukawa2007Topological} remains largely unexplored due to significant computational challenges.

Based on the description above, it can be seen that previous studies employing ED and QMC methods have demonstrated significant difficulties due to the frustrations and constraints in the QDM model. ED works only for small clusters. QMC simulations, on the other hand, can be quite challenging, especially when addressing the highly degenerate states that encompass all maximally flippable plaquette configurations. Previous works based on the zero-temperature Green's function QMC~\cite{Ralko2005,Ralko2006,Vernay2006Identification,Ralko2007}
and path-integral-based sweeping cluster QMC~\cite{YanTopological2021} on triangular lattice QDM, require extensive sampling and either very long projection or very low temperatures to obtain meaningful results in columnar, $\sqrt{12}\times\sqrt{12}$ VBS and the $\mathbb{Z}_2$ QSL phases. The computational cost scales rapidly with system size, and the criticality of the VBS-QSL transition is still not fully determined (see below).

It is under such circumstances, we find it interesting that the matrix product state (MPS) or tensor-based methods have not been fully applied to constrained lattice models. To the best of our knowledge, previous density matrix renormalization group (DMRG) attempts for constrained models have been done by projecting to the good basis states that satisfy the constraint at each level of the algorithm and filtering out the states that do not satisfy the QDM constraints~\cite{chepiga_floating_2019,chepigaDMRG2019,Chepiga2024z4}; by adding additional potential terms to penalize states that violates the dimer constraints~\cite{Lee2016Electronic}; or by introducing redundancy in local physical space to take care of the constraint and having local gauge symmetry to compensate the redundancy~\cite{Tschirsich_phase_2019}. All practices work well when the system is 1D-like (e.g., a ladder or cylinder with small circumference) or has a simpler lattice structure (square lattices), but will become cumbersome when approaching the 2D limit and for models with constraints that are hard to cure by local redundancy. The last approach introduced the idea of including local symmetry to avoid projection or penalty terms, following the common practice in spin and fermion models with global symmetry such as $\mathrm{U}(1)$ or $\mathrm{SU}(2)$ symmetries~\cite{White_DMRG_1992,White_DMRG_1993, McCulloch_DMRG_2007, SCHOLLWOCK_DMRG_2011, Singh_Tensor_2011, ORUS_practical_2014}.
%The important property of MPS and matrix product operator (MPO) in DMRG, that is, the symmetry of the model, has not been exploited in previous attempts for the QDM, certainly not to the level of the symmetry implementation in the MPS and MPO for spin and fermion models with only global U(1) or SU(2) symmetries~\cite{White_DMRG_1992,White_DMRG_1993, McCulloch_DMRG_2007, SCHOLLWOCK_DMRG_2011, Singh_Tensor_2011, ORUS_practical_2014}.
The direct conversion of the local constraint into symmetry implementation is where our development is built upon.

In this work, we introduce a new DMRG design to study systems with local constraints, and we successfully apply it to the triangular QDM on cylinder geometries. Our approach {\it symmetrizes the constraints}, in that it treats the constraint as a set of local symmetries and implements them directly within the DMRG framework to block diagonalize all local tensors. This method inherently satisfies the constraints and significantly accelerates the computation, making it applicable to any 2D system with local constraints, with the same computational complexity of DMRG for 2D systems with global symmetries. By combining this method with ED analysis, we study the ground-state phase diagram of the QDM on the triangular lattice. Our results with wide cylinders ($L_y=4,6,8$) successfully capture the columnar solid phase, $\sqrt{12}\times \sqrt{12}$ valence bond solid phase and the $\mathbb{Z}_2$ quantum spin liquid phase, at different parameter regions of the ground state phase diagram and are fully consistent with previous works on the problem with other methods~\cite{moessnerIsing2001,Moessner2001b, moessnerShort2001,moessnerResonating2001,YanTopological2021,Ralko2005,Ralko2006,Ralko2007}. Furthermore, we investigate the EE in both the $\mathbb{Z}_2$ topological QSL phase, and for the first time, reveal the topological EE of $\gamma=-\ln(2)$ inside the QSL phase at parameters deviated from the RK point. These results are beyond the existing literature and point out the great potential of our DMRG algorithm in studying other constrained quantum many-body systems such as quantum loop models and synthetic quantum simulators.

The rest of the paper is organized as follows. In Sec.~\ref{sec:II}, the main idea of our method is given, focusing on the implementation of the {\it symmetrization of local constraints} and {\it dynamic storage of quantum numbers}. Sec.~\ref{sec:III} discusses our results on triangular lattice QDM, where Sec.~\ref{sec:IIIA} shows the benchmark with ED for exact results and Sec.~\ref{sec:IIIB} gives the structure factor of the dimer correlation function based on wide cylinders that exhibit the columnar phase, the $\sqrt{12}\times\sqrt{12}$ phase and the $\mathbb{Z}_2$ QSL phase in the ground state phase diagram of the model, and Sec.~\ref{sec:IIIC}, on the other hand, focuses on the EE results from our DMRG and for the first time in literature, we show the topological EE inside the $\mathbb{Z}_2$ QSL away from the exactly solvable RK point. Finally, Sec.~\ref{sec:IV} reiterates the reason for the success in our symmetrization of the local constraints and points out other interesting constrained quantum lattice models that are either of fundamental theoretical importance or of immediate experimental relevance in various quantum simulators that are waiting to be explored with our algorithm.

\begin{figure*}[htp!]
	\centering
	\includegraphics[width=\textwidth]{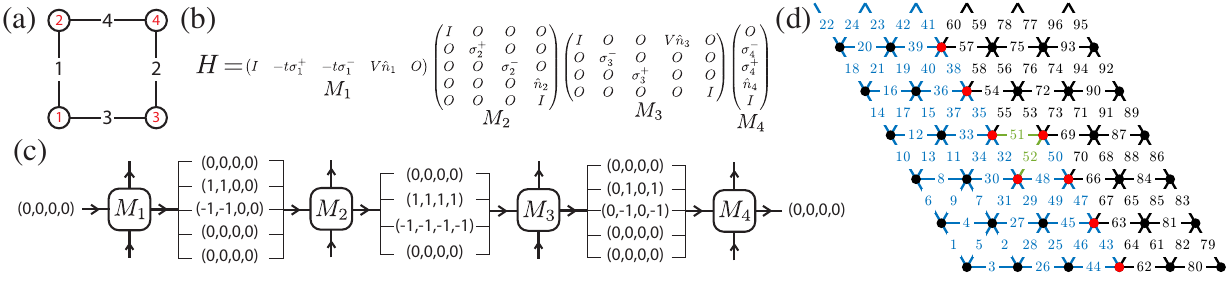}
	\caption{\textbf{QDM on a plaquette as an algorithmic demonstration.} (a) The system with black (red) numbers indexing the bonds (sites). (b) The Hamiltonian in terms of local tensors. (c) Schematic of the quantum numbers flows for the MPO in (b). The four numbers are the quantum numbers (number of dimers on each site), and each four numbers set label a virtual bond space. (d) A $6 \times 6$ triangular lattice with OBC (PBC) along $x$- ($y$-) direction. The local degree of freedom lives on every bond, and they are labeled according to the DMRG path. For a two-site update on bonds 51 \& 52 (green), only 8 quantum numbers (sites in red) are active. All bonds before 51 are colored in blue.}
	\label{fig:fig_MPO}
\end{figure*}

\section{Method}
\label{sec:II}

We consider the Hamiltonian of QDM on a triangular lattice, 
\begin{eqnarray}
  H=&-t&\sum_\alpha \left(
  \left|\rhombV\right>\left<\rhombH\right| + \left|\rhombH\right>\left<\rhombV\right|
  \right) \nonumber \\
  &+V&\sum_\alpha\left(
  \left|\rhombV\right>\left<\rhombV\right|+\left|\rhombH\right>\left<\rhombH\right|
  \right),
\label{eq:hamiltonian}
\end{eqnarray}
where the $t$-term flips parallel dimers to other directions in the plaquette and the positive (negative) $V$-term repulses (attracts) parallel dimers. The RK point located at $t=V$ is an exact-solvable point, where the ground state wavefunction consists of the equal superposition of all possible dimer configurations under the constraint~\cite{rokhsarSuperconductivity1988,moessner2010quantum}.

Note that every term in the Hamiltonian commutes with the operator counting the number of dimers connected to each site. Although each kinetic ($t$) term swaps dimer on bonds with different orientations, the four sites it touches are connected to the same numbers of dimers before and after. Consequently, the number of dimer connected to each site is a conserved quantity, i.e., the local constraint, and we have $N_\textrm{site}$ many of them defined on the lattice. One can then implement these conserved quantities in the DMRG computation, the same as how one considers the total $S^z$ in the Heisenberg model as a good quantum number. Once these symmetries are implemented, one can speed up all computation by making use of the sparse block structure of local tensors in MPO and MPS, and the constraint will be set by the initial MPS and unchanged throughout the computation.

Moreover, the Hamiltonian commutes with operator $M=\prod_{i\in l_M}(-1)^{\hat{n}_i}$ leading to two topological sectors, where $l_M$ is one horizontal path cutting through the whole system (red dashed line in Fig.~\ref{fig:fig_benchmark} (b) \& (c)). The commutation relation follows from the fact that every plaquette operator in the Hamiltonian touches either 0 or 2 bonds on $l_M$, leaving $M$ unchanged. Since we only have PBC along the $y$-direction, this geometry holds only two topological sectors characterized by $M=\pm1$. This global symmetry can also be implemented in DMRG by tracking a quantity that changes sign upon crossing a dimer located on bonds along $l_M$.

Before delving into the details of implementing local symmetries, let's begin with an example of the Heisenberg model. The Hamiltonian
\begin{equation}
H=J\sum_{\langle i,j\rangle}\vec{S}_i\cdot \vec{S}_j=J\sum_{\langle i,j\rangle}\{S^z_iS^z_j+\frac{1}{2}(S^+_iS^-_j+S^-_iS^+_j)\}
\label{eq:eq2}
\end{equation}
The Hamiltonian possesses global $\mathrm{SU}(2)$ and $\mathrm{U}(1)$ symmetry. The latter is related to the conservation of the total $S^z=\sum_i S^z_i$, meaning $[H,S^z]=0$. To implement this symmetry in DMRG, one need to (1) mark all local tensors in MPS and MPO with proper charge flow, (2) perform singular value decomposition (SVD) to each block labeled by the same charges flowing in and out when SVD is required on the full tensor, and (3) start the DMRG process with an initial MPS, which determines the charge number of the state and set the charge sector for this simulation.

In an MPO, the charge flow is marked when constructing the Hamiltonian MPO in the automata process~\cite{White_DMRG_1992,White_DMRG_1993, McCulloch_DMRG_2007, SCHOLLWOCK_DMRG_2011, Singh_Tensor_2011, ORUS_practical_2014}. Suppose one would like to put in the interaction $S^+_2 S^-_3$, where 2 and 3 are two nearest neighbors in Eq.~\eqref{eq:eq2}. In the second local tensor in the MPO train, an $S^+_2$ is put in a block labeled 0 and +1 for charges flowing in and out in the virtual bond spaces, respectively, and $S^-_3$ is put in the third local tensor, with labels +1 and 0. Since all the terms in the Hamiltonian preserve such symmetry, every term will end with charge 0. By labeling flows for all dimensions like this, all local tensors are block diagonalized, where nonzero elements only appear in blocks where charges flowing in and out are the same, thus, SVD can be performed on each block successively. For the MPS, the charge flow in each local tensor is marked similarly, and the charge flowing out from the last tensor is the total $S^z$ for the state.

Our DMRG algorithm in QDM is essentially the same, except that we do not have a global symmetry. Instead, we have $N_\mathrm{site}$ local symmetries, corresponding to the conservation of the number of dimers on every site. And instead of labeling flows with one quantum number, we need $N_\mathrm{site}$ of them. Such an implementation in DMRG is the new development in our method.

For QDM or QLM, as mentioned before, the Hamiltonian preserves the number of dimers connected to each site.  Therefore, we can label the MPO and MPS similarly to how it is done in the Heisenberg model, with each virtual bond space being labeled with $N_\mathrm{site}$ quantum numbers. For the last tensor of MPO, the quantum numbers will flow to $(0,0,\cdots,0)$, similar to the Heisenberg case. For the MPSs, the last flowing out should represent the number of dimers connected to each site in the state, which should be $(1,1,\cdots,1)$ for QDM and $(2,2,\cdots,2)$  for QLM, respectively.

Consider the simplest case involving only 1 plaquette (Fig.~\ref{fig:fig_MPO} (a)). The Hamiltonian is simply
\begin{equation}
H=-t (\sigma_1^+\sigma_2^+\sigma_3^-\sigma_4^- + \sigma_1^-\sigma_2^-\sigma_3^+\sigma_4^+)+ V(\hat{n}_1\hat{n}_2+\hat{n}_3\hat{n}_4), 
\label{eq:eq3}
\end{equation}
where Pauli matrices $\sigma_i^+$, $\sigma^-_i$ acting on the two-dimensional local Hilbert space on the $i$-th bond creates and annihilates a dimer respectively, and $\hat{n}_i$ counts the number of dimers on the $i$-th bond. The MPO train formulation for this Hamiltonian is shown in Fig.~\ref{fig:fig_MPO} (b). One can recover the Hamiltonian by contracting all virtual bond dimensions.

%%%%%%%%%%%%%%%%%%%%%%%%%%%%%%
% $$
% \begin{pmatrix}
% I & -t \sigma^+_1 & -t \sigma^-_1 & V\hat{n}_1 & O
% \end{pmatrix}
% $$
% $$
% \begin{pmatrix}
% I & O & O & O\\
% O & \sigma^+_2 & O & O \\
% O & O & \sigma^-_2 & O \\
% O & O & O & \hat{n}_2 \\
% O & O & O & I
% \end{pmatrix}
% $$
% $$
% \begin{pmatrix}
% I & O & O & V \hat{n}_3 & O\\
% O & \sigma^-_3 & O & O & O \\
% O & O & \sigma^+_3 & O & O \\
% O & O & O & O & I
% \end{pmatrix}
% $$
% $$
% \begin{pmatrix}
% O \\ \sigma^-_4 \\ \sigma^+_4 \\ \hat{n}_4 \\ I
% \end{pmatrix}
% $$
%%%%%%%%%%%%%%%%%%%%%%%%%%%%%%

\begin{figure*}[htp!]
	\centering
	\includegraphics[width=\textwidth]{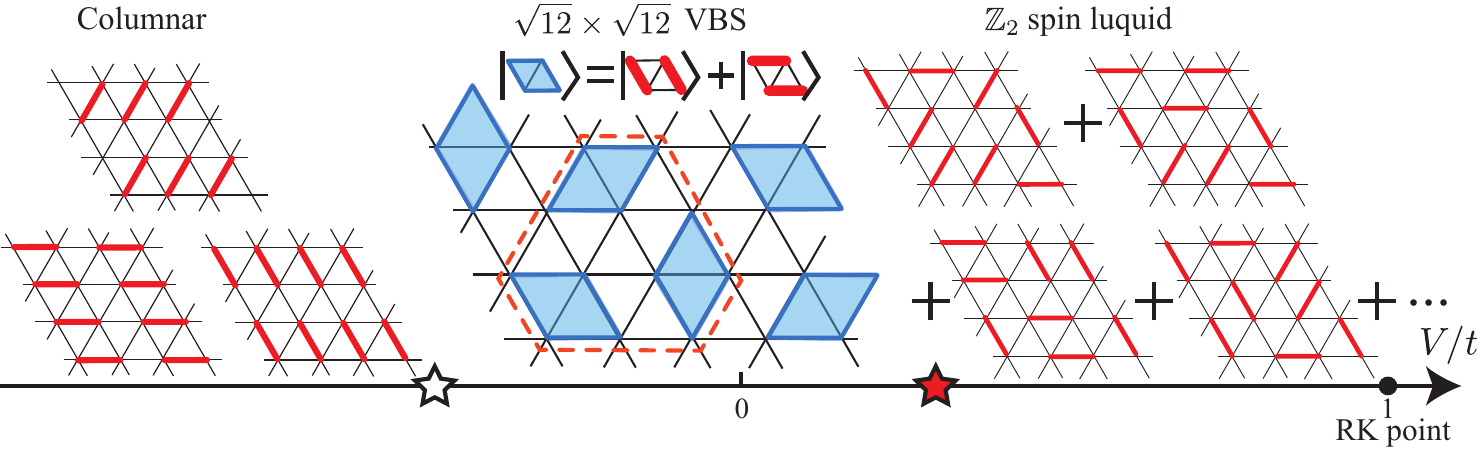}
	\caption{\textbf{Expected Ground State Phase Diagram.} For negative values of $V$, the diagonal term that attracts dimers becomes dominant, leading the system to form parallel columns along one of the three directions. As $V$ increases to approximately  $-0.7$~\cite{Ralko2005}, the hopping term induces resonating rhombi tiling across the system, forming the $\sqrt{12}\times\sqrt{12}$ VBS pattern. When $V$ continues to increase to around $0.8$~\cite{Ralko2006,Ralko2007,YanTopological2021}, the system transitions into a highly quantum-entangled spin liquid phase extended to the RK point at $V=1$.}
	\label{fig:fig_phase_diagram}
\end{figure*}

Consider the first term in the Hamiltonian $-t \sigma_1^+\sigma_2^+\sigma_3^-\sigma_4^-$, which is carried in the second channels (columns) in local tensors $M_1$,$M_2$ and $M_3$. $-t \sigma_1^+$ acting on the first bond creates a dimer here, thus increase the quantum numbers (numbers of dimer) on sites 1 and 2 (Fig.~\ref{fig:fig_MPO} (a)) by one, thus leads to quantum numbers flowing out at the second channel being (1,1,0,0). In $M_2$, $\sigma_2^+$ is placed at the (2,2) block, which means the quantum numbers flowing in is (1,1,0,0) (the quantum numbers flowing out of the second channel in $M_1$). A creation on bond 2 will increase quantum numbers on sites 3 and 4 by 1, thus leading to quantum numbers flowing out at the second channel being (1,1,1,1). One can follow any of the terms and check that all quantum numbers indeed flow to (0,0,0,0) eventually (Fig.~\ref{fig:fig_MPO} (c)), as guaranteed by the symmetries of Hamiltonian. More details on the implementation can be found in Appendix~\ref{sec:DMRG_detail}.

In a larger system with more sites or a more complicated lattice structure, the same symmetry can be implemented. However, in practice, one does not need to keep track of all $N_\textrm{site}$ many quantum numbers for the whole process, but only the active ones, which are the ones touched by previous bonds at least once and will be touched later.  Consider a $6\times6$ triangular lattice, as shown in Fig.~\ref{fig:fig_MPO} (d), where the numbers indicate the sequence of DMRG sweep, and blue (black) bonds represent the visited (unvisited) bonds. If a two-site update is performed on local tensors 51 \& 52 (bonds in green), only 8 quantum numbers (sites in red) are active. For sites connected to only blue bonds, their corresponding quantum numbers cannot be modified anymore and are equal to the values set by the constraint, which are the same throughout all virtual bond dimensions. For sites connected to only black bonds, they have yet to be touched by any dimer, and their corresponding quantum numbers are all 0. Only those sites on the boundary of blue and black bonds carry active quantum numbers.
% All the others are either finished before bond 51 and cannot be modified anymore or are yet to be touched until after bond 52, and the corresponding quantum numbers are all 0. 
For DMRG simulation on a 2D lattice wrapped around a cylinder, the number of such non-trivial quantum numbers to be stored at each local tensor can be reduced to roughly the number of sites along the periodic boundary condition ($L_y$).

As a result, the memory required to label each local tensor scales as $\chi L_y$ rather than $\chi N_\textrm{site}$, thereby restoring the typical linear scaling with $L_x$ observed in DMRG.

In the worst-case scenario, the number of all possible quantum number combinations scales as 2 to the power of the number of active quantum numbers, each taking the value 0 or 1. Accordingly, the bond dimension must be large enough to accommodate all possible states, implying that it should scale exponentially with $L_y$—as is typical in standard DMRG applied to 2D systems wrapped around a cylinder. In practice, however, this number is substantially reduced due to local constraints that prevent the quantum numbers from being entirely independent.

To summarize, the key ingredients of our symmetrization of the local constraint DMRG method for QDM are  
\begin{enumerate}
  \item \emph{Symmetrization of local constraint}---Tracking the flow of $N_\textrm{site}$ conserved quantities, that is, the numbers of dimers on each site, thus automatically fulfilling the constraint, and
  \item \emph{Dynamic storage of quantum numbers}---Carrying and utilizing only a small amount of the quantum numbers that are ``active'' during the DMRG process.
\end{enumerate}

With such implementation, we found the DMRG simulation can be readily applied on the QDM/QLM on 2D lattice models, with the same computational complexity as those on Heisenberg or Hubbard type models. Below, we will use the triangular lattice QDM as an example to demonstrate the power of our algorithm.

\section{Results}
\label{sec:III}
\begin{figure}[htp!]
	\centering
	\includegraphics[width=\columnwidth]{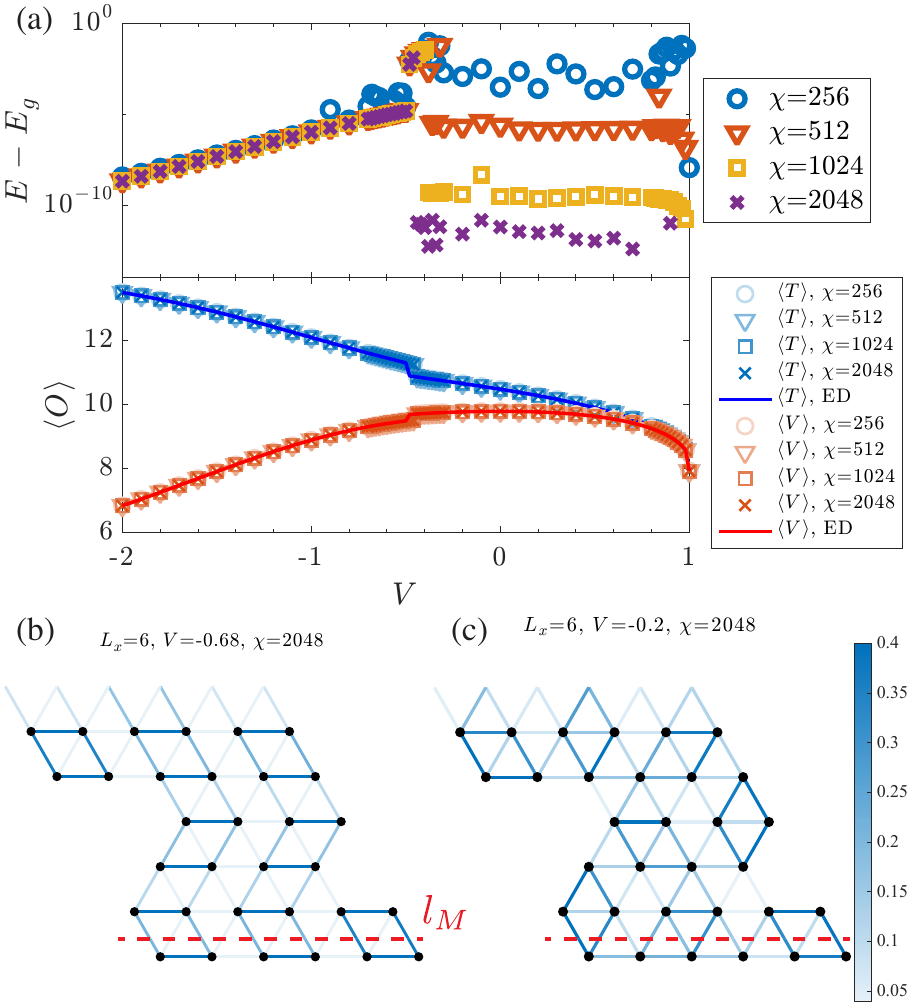}
	\caption{\textbf{Results for $L_x=L_y=6$ and benchmark with ED.} (a) Energy difference (with $E_g$ the ground state energy from ED) and operator expectation values obtained from ED and DMRG. Inside the columnar phase ($V<-0.5$), the DMRG has difficulty in resolving the 2 degenerate ground states and accuqire a difference of $10^{-5}$ to $10^{-10}$ compared with ED, the difference becomes smaller as $V$ moves deep into the columnar phase. Inside the VBS and QSL phases ($V>-0.5$), the energy difference between DMRG and ED is at machine precision as bond dimension increases. Expectation values of the physical observables, $\langle T\rangle$ and $\langle V \rangle$ terms of the Hamiltonian, have no noticeable difference throughout the $V$ range we investigated. (b) and (c) are the dimer density of $V=-0.68$ (inside the columnar phase) and $V=-0.2$ (inside the VBS phase) from DMRG. One sees the columnar arrangement of the dimers in the former and the $\sqrt{12}\times\sqrt{12}$ unit cell in the latter.}
	\label{fig:fig_benchmark}
\end{figure}

In Secs.~\ref{sec:IIIA} and \ref{sec:IIIB}, we present the DMRG result for the quantum dimer model on the triangular lattice warp around a cylinder with fixed circumference $L_y=6$, with periodic (open) boundary condition along the $y$ ($x$) direction. To satisfy the columnar and $\sqrt{12}\times\sqrt{12}$ VBS phases (see Fig.~\ref{fig:fig_phase_diagram}) with such boundaries under the 1 dimer per site condition, the $L_y$ has to be multiples of 6 and a few sites near the boundary is discarded (see Fig.~\ref{fig:fig_benchmark} (b) \& (c) for the geometry in the $6\times 6$ case). In Sec.~\ref{sec:IIIC}, we present the results of the EE computation with $L_y=2,4,6$.

The ground state phase diagram of the model, Fig.~\ref{fig:fig_phase_diagram}, as a function of $V$ (set $t=1$ as the energy unit) has been intensively investigated with different techniques, ranging from field theory to ED, (variational) Green's function Monte Carlo (GFMC), and sweeping cluster QMC~\cite{moessnerResonating2001, Ralko2005, Ralko2006, Ralko2007, YanTopological2021}. The ED and GFMC studies, which involve extrapolations of the level crossing of the topological gap as well as the decay of the dimer and vison gaps, indicate that the transition point between the the columnar and $\sqrt{12}\times\sqrt{12}$ VBS phase occurs at $V\sim -0.7$~\cite{Ralko2005}, and the transition point between the $\sqrt{12}\times\sqrt{12}$ phase and the $\mathbb{Z}_2$ QSL is estimated to occur at $V\sim 0.8$~\cite{Ralko2006, Ralko2007}. Furthermore, sweeping cluster QMC results suggest consistent points of the transition point between the $\sqrt{12}\times\sqrt{12}$ phase % and the $\mathbb{Z}_2$ QSL is $V=0.85(5)$
, derived from the analysis of dimer and vison spectra~\cite{YanTopological2021}. Notably, the precise position of these transition points, with statistical error bars, have not been determined due to the computation difficulties mentioned above, and there have been no prior DMRG studies reported on this system. 

It is expected that at negative $V$, the dimers will form a columnar phase with translational symmetry breaking in one of the three lattice directions (related by 120$^{\circ}$ rotation) and 2-fold degeneracy in each of the symmetry-breaking directions. The formation of the columnar phase is due to the interplay of the attractive $V$ and an order-by-disorder effect~\cite{moessnerMagnets2001,moessnerIsing2001,moessnerResonating2001,zhitomirskyField2002} (to overcome the extensive degeneracy $\exp(L_y)$ generated by sliding each row of dimers at the ordered direction freely). As $V$ moves towards positive values, the columnar phase will transit into a valence bond solid (VBS) with $\sqrt{12}\times\sqrt{12}$ unit cell~\cite{moessnerResonating2001}. Such an enlarged unit cell contains 3 resonating rhombi, i.e., each unit cell contains 12 lattice sites, and it is supposed to have a Bragg peak of the dimer-dimer correlation at the $\vec{X}$ point of the Brillouin zone (BZ). The transition between columnar and VBS phase is suggested to be first order~\cite{moessnerResonating2001, Ralko2005}.

Further increase $V$ towards the RK point of $V=t=1$~\cite{rokhsarSuperconductivity1988}. The VBS phase is expected to go through a continuous transition to a $\mathbb{Z}_2$ QSL phase~\cite{moessnerResonating2001,moessnerIsing2001,Ivanov2004,Ralko2005,Ralko2006,Ralko2007,KrishanuTopological2015,YanTopological2021}. Since the lattice is non-bipartite, the RK point is also inside the QSL phase with gapped vison excitations~\cite{YanTopological2021}, the VBS-QSL transition, can therefore be interpreted as the condensation of the visons, to confine these fractionalized excitations into the dimer (a confined pair of vison) and establish the dimer order of the VBS phase, with only the short-range resonating of the dimers within the $\sqrt{12}\times\sqrt{12}$ unit cell as the remnants of the fractionalization. And since the vison dispersion has 4 degenerate momenta that are going to close gap at the transition~\cite{moessnerIsing2001}, it was suggested that the critical theory of this transition belongs to the (2+1)D O(4)* universality~\cite{moessnerIsing2001,YanTopological2021}, where the continuous symmetry O(4) is emergent and the * means that the transition separate symmetry-breaking VBS and a symmetric and yet topologically non-trivial $\mathbb{Z}_2$ QSL with fractionalized visons. Similar * type (2+1)D quantum critical points (QCP), have been well studied in the frustrated quantum spin models with (2+1)D XY* transition~\cite{sachdev_book,Isakov2012,YCWang2018}, and more recently in the triangular lattice QLM with (2+1)D Cubic* transition, separating a vison VBS phase and $\mathbb{Z}_2$ QSL~\cite{RanHidden2024,RanCubic2024,RanPhase2024}. 

As mentioned above, the precise positions of the columnar-VBS first-order transition and the VBS-QSL O(4)* transition are less accurately determined due to the challenges associated with computing constrained quantum lattice models and extrapolating to the thermodynamic limit, despite previous intensive studies. Moreover, the verification of the O(4)* has not been carried out using controlled numerical simulations. It is expected that the former is $V\sim -0.7$~\cite{Ralko2005} and the latter is $V\sim 0.8$~\cite{Ralko2006,Ralko2007,YanTopological2021}.
As we will show later, our DMRG results indeed find the consistent phases according to these previously determined boundary values.
%are consistent with such previous expectations. 
Admittedly, in finite-size simulations, the precise boundary value might differ due to the finite size effect (the situation here is more complicated as the thermodynamic limit values have not been precisely determined). Previous ED studies of small systems suggested that $V=-1$ is still within the VBS phase for $N=48$ cluster~\cite{Wietek2024Quantum}; we take the cautious attitude here that we only discuss the computed properties inside each phase rather than determining the transition points.

The ground state mainly resides in the trivial topological sector ($M=+1$), and in $M=-1$ only within the VBS phase when $L_x/2$ is odd. For the following example with $L_x=6$, the columnar (VBS) state belongs to the $M=+1\;(-1)$ sector. In the case of the QSL state, the two topological sectors are degenerate.

\begin{figure*}[htp!]
	\centering
	\includegraphics[width=\textwidth]{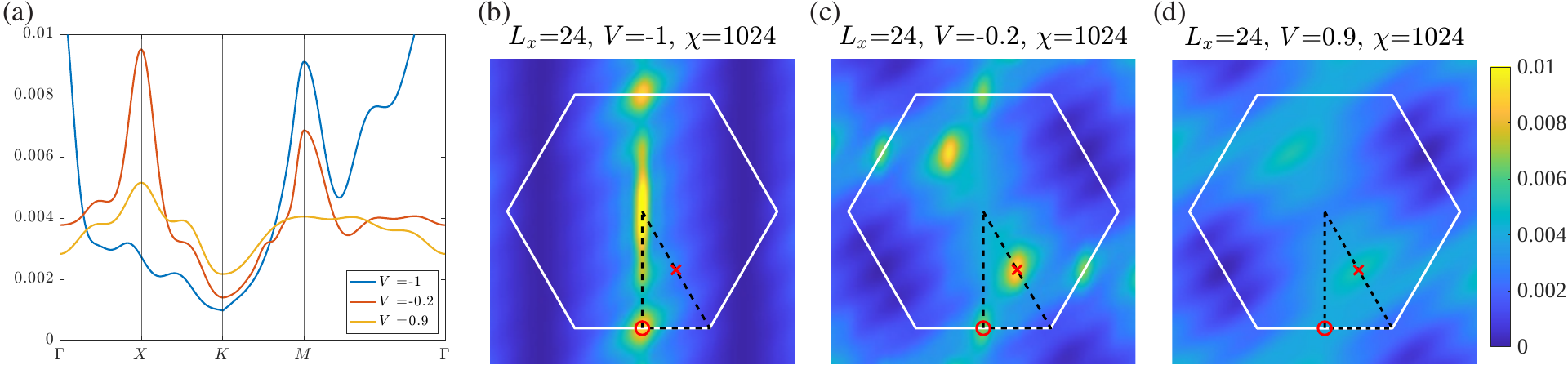}
	\caption{\textbf{Dimer correlation of QDM on triangular lattice.} (a) Dimer correlation function $S(\vec{k})$ along the high symmetry path (denoted as the black path in panels (b-d)) for $V=-1$, $V=-0.2$ and $V=0.9$. The system size is $L_x=24$ and $L_y=6$. (b-d) are the correlation in the entire BZ. Inside the columnar phase with $V=-1$, the Bragg peak of the order is at the $\vec{M}$ points (red circle); inside the $\sqrt{12}\times\sqrt{12}$ VBS phase with $V=-0.2$, the order wave vector is at $\vec{X}$ points (red corss), and the $\vec{M}$ point becomes secondary; inside the $\mathbb{Z}_2$ QSL phase with $V=0.9$, the dimer correlation is flat as the dimer (as well as vison) correlation are all short-ranged in real-space~\cite{YanTopological2021}. Panels (b-d) share the same colorbar on the right.}
	\label{fig:fig_result}
\end{figure*}

\subsection{Benchmark with exact diagonalization}
\label{sec:IIIA}
For the smallest system $L_x=L_y=6$, we performed exact diagonalization (ED) to serve as a benchmark for DMRG. The ED is carried out by selecting states satisfying the one dimer per site criteria; one then constructs and diagonalizes the Hamiltonian in this restricted Hilbert space.

Fig.~\ref{fig:fig_benchmark} shows the DMRG result compared to that from ED. Panel (a) shows the energy difference and operator expectation values obtained from the two methods. For negative $V$, the system is in a columnar phase, and the ground state is (almost) two-fold degenerate with a finite size gap of order $10^{-9}\sim10^{-5}$. The DMRG process is not able to separate the two states and thus leads to a discrepancy in energy compared with that of the ED $E_g$ inside the columnar phase ($V<-0.5$) for all bond dimensions. 

As $V$ moves toward positive values of the VBS and QSL phases, our finite size system is non-degenerate and gapped, and the DMRG can find the ground state when enlarging bond dimension. For $V>-0.5$, as we increase the DMRG bond dimension $\chi$, the energy difference between DMRG and ED, $E-E_g$, goes towards the machine precision $10^{-15}$. 

The lower half of Fig.~\ref{fig:fig_benchmark} (a) shows the expectation values with two parts of the Hamiltonian ($t$-term and $V$-term), respectively, obtained from the inner product of the operators and ground states in each method ($\langle O\rangle=\langle\psi |O|\psi\rangle$). DMRG results (markers) align well with the ED results (solid lines) for all $V$. It is interesting to observe that even on such a small size, the columnar first-order transition manifests around $V=-0.5$.

Panels (b) and (c) of Fig.~\ref{fig:fig_benchmark} show the dimer density for two different $V$, which presents the characteristic real space patterns of columnar and VBS phases, respectively.

\subsection{Ground state phase diagram}
\label{sec:IIIB}

Fig.~\ref{fig:fig_result} shows our results on the ground state phase diagram of the model. Here, we summarize data with $L_x=24$ and $L_y=6$, with the same open boundary condition in the $L_x$ direction as in Fig.~\ref{fig:fig_benchmark}, the bond dimension $\chi=1024$ has been tested to be converged for this system size.

We show the obtained dimer-dimer correlation function $S(\vec{k})=\frac{1}{N^2}\sum_{i,j}e^{-i\vec{k}\cdot (\vec{r}_i-\vec{r}_j)}\bra{\psi}(\hat{n}_i-\frac{1}{6})(\hat{n}_j-\frac{1}{6})\ket{\psi}$ in the Brillouin zone at $V=-1$ (inside the columnar phase), $V=-0.2$ (inside the VBS phase) and $V=0.9$ (inside the QSL phase). The OBC along $x$-direction breaks the translational symmetry, rendering $k_x$ not a good quantum number. However, our computation of $S(\vec{k})$ focused solely on the correlations in the bulk region, which is sufficiently far from the boundary, where the translational symmetry remains largely intact. In Fig.~\ref{fig:fig_result} (a) and (b), it is clear that the Bragg peaks are at one of three $\vec{M}$ points, as we chose only bonds towards the top left direction in this calculation. As $V$ moves to $-0.2$ (Fig.~\ref{fig:fig_result} (a) and (c)), the Bragg peak moves to the $\vec{X}$ point, which according to the previous literature~\cite{Ralko2006}, is the wave vector of the $\sqrt{12}\times\sqrt{12}$ order from dimer correlation function. Here, again, we observe one of the three pairs of the $\vec{X}$ peaks, and the other two can be seen if one looks at bonds in other directions. It is interesting to observe that inside the VBS phase, the Bragg peak at the $\vec{M}$ point is also present, although it is secondary. This is also consistent with the results in previous QMC works~\cite{Ralko2006,Ralko2007,YanTopological2021}. As $V$ moves inside the QSL phase (Fig.~\ref{fig:fig_result} (a) and (d)), the dimer correlation in the entire BZ flattens, as there is no long-range order in the dimer arrangement and both dimer and vison spectra are gapped~\cite{YanTopological2021}. The system acquires with $\mathbb{Z}_2$ topological order and one can detect such long-range entanglement with the scaling of the entanglement entropy~\cite{kitaevTopological2006,levinDetecting2006,zhaoMeasuring2022,chenTopological2022}, as we now turn to. The results of the order parameters and entanglement entropy scanning through the ground state phase diagram can be found in Appendix~\ref{sec:Ground_state_result}.

\subsection{Entanglement entropy inside the $\mathbb{Z}_2$ QSL phase}
\label{sec:IIIC}

For a system, we can bipartite it into subregions $A$ and $B$, and the reduced density matrix can be obtained by partial tracing out the degree of freedom in subregion $B$. Specifically, we have $\rho_A=\sum_{\xi}\bra{\xi}\rho\ket{\xi}$, where $\rho=\ket{\psi}\bra{\psi}$ is the density matrix for the entire system and $\{\ket{\xi}\}$ forms a complete basis in $B$. The entanglement entropy is defined as
$S^\textrm{vN}=-\Tr{}\;\rho_A\ln\rho_A$
characterizes the quantum entanglement between the two subregions. The scaling behavior of this entropy offers insights into various quantum phases.

For the $\mathbb{Z}_2$ QSL, the total quantum dimension ${\mathcal{D}=\sqrt{1^2+1^2+1^2+1^2}=2}$, corresponds to the four types of anyons in the $\mathbb{Z}_2$ gauge theory, leading to topological entanglement entropy (TEE) $\gamma=\ln(\mathcal{D})=\ln(2)$ on the cylindrical geometry~\cite{kitaevTopological2006,levinDetecting2006,Jiang_Identifying_2012}. With a smooth boundary of length $l$, the bipartite entanglement entropy scales as
\begin{equation}
    S= a l - \gamma,
\end{equation}
with non-universal leading term coefficient $a$. We note, the determination of the TEE for other simpler realization of the $\mathbb{Z}_2$ (also $\mathbb{Z}_3)$ topological order either in the toric code model, frustrated spin and hard-core boson models, have only been achieved in DMRG or QMC simulations~\cite{isakovTopological2011,Jiang_Identifying_2012,zhaoMeasuring2022,chenTopological2022}, none of which possesses local constraints such as that in QDM. 

The difficulty of computing the EE in QDM again comes from the difficulty in handling the local constraint and yet still being able to extrapolate to larger system sizes. There were previous ED and classical Monte Carlo simulation (where the closed-form wavefunction is know) at the RK point~\cite{Furukawa2007Topological,selemEntanglement2013,peiRenyi2014}. And the expected $\gamma$ for $\mathbb{Z}_2$ topological order on triangular lattice QDM at the RK has been obtained. But the attempt away from the RK point with a small size ED cannot yield the correct $\gamma$~\cite{Furukawa2007Topological} and therefore, the knowledge and understanding of the $\mathbb{Z}_2$ topological order inside the QSL phase, from a quantum entanglement point of view, is still missing.

% Moreover, on a finite size system, the EE can also reveal the degeneracy of the ordered ground state. For example, it is expected on a finite size lattice
% \begin{equation}
%     S= a l + d,
% \end{equation}
% where $d$ is the logarithm of the number of ground state degeneracy due to the symmetry breaking~\cite{PlatMagnetization2015}. One can also test this behavior on the cylinder geometry inside the columnar phase, where $d=\ln(2)$ is expected as only one of the three columnar directions is favored.

In our DMRG simulation of cylindrical geometry, the spin liquid phase is two-fold degenerate, one from each topological sector ($M=\pm1$). Previous studies revealed that the constant term in the scaling of EE differs within the two-dimensional ground state eigen-space when the bipartition boundary is non-contractable~\cite{Zhang_Quasiparticle_2012,Jiang_Identifying_2012,zhaoMeasuring2022}. Luckily for DMRG, the computation finds the optimized state with minimal EE (maximal $\gamma$). And such minimal entropy state (MES) is a combination of states from two sectors, giving $\langle M\rangle=0$. Namely, one needs to perform DMRG \emph{without} separating the two topological sectors.

\begin{figure}[htp!]
	\centering
	\includegraphics[width=\columnwidth]{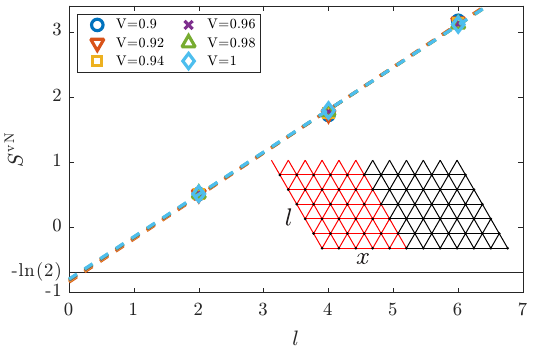}
	\caption{\textbf{Extrapolation of EE in $\mathbb{Z}_2$ quantum spin liquid and columnar phases.} With a cylinder geometry and smooth boundary, the TEE (the intercept of the dashed fitting line) extrapolates to $-\ln(2)$ using data with boundary length $l=L_y=2,4,6$, not only at the RK point with $V=1$ but also inside the QSL phase with $V=0.9,0.92,0.94,0.96$ and 0.98. The inset shows an example of a smooth cut, where the red bonds are considered as subregion $A$ and black bonds as $B$.}
	\label{fig:fig_EE_scaling}
\end{figure}

Fig.~\ref{fig:fig_EE_scaling} shows the result of von-Neumann entanglement entropy $S^\textrm{vN}$ for a few $V$, where the system is believed to be in the $\mathbb{Z}_2$ spin liquid phase. We reverted to the smooth boundary in this computation since we do not need to satisfy the VBS phase here. We compute the EE with $L_y=2,4,6$ and $L_x$ is kept to be 24, where EE has stabilized and can reflect bulk information. The system is cut in half along the y direction, creating a smooth boundary with length $l=L_y$. Using only the three smallest possible circumferences, we are able to obtain intersections around 0.7(1) by linear fitting, which is close to the anticipated $\gamma=\ln(2)\approx0.693$. We note that such precise determination of the TEE at the thermodynamic limit, inside the $\mathbb{Z}_2$ QSL phase away from the RK point, is reported for the first time.

\section{Discussion}
\label{sec:IV}
With all the results on triangular QDM presented here, we are delighted that our DMRG method can indeed study systems with local constraints beyond the 1D and ladder settings. Different from the previous attempt to implement DMRG for such constrained models, where projecting MPS and MPO back into the subspace to satisfy the constraint has to be applied~\cite{chepiga_floating_2019,chepigaDMRG2019,Chepiga2024z4}, our approach naturally treats the constraint as a set of local symmetries, and implement them directly within the DMRG framework to speed up the computation. This method inherently satisfies the constraints and significantly accelerates the computation, making the generic 2D system with local constraints as simulatable as the 2D Heisenberg or Hubbard models with DMRG. We have therefore obtained the characteristic information of the columnar, $\sqrt{12}\times\sqrt{12}$ VBS and $\mathbb{Z}_2$ QSL phases, and for the first time, the entanglement fingerprint of the topological order away from the special RK point.

We believe the method developed here open the new opportunities for matrix and tensor based algorithms for constrainted quantum lattice models, such as the QDM/QLM on different geometries~\cite{KrishanuTopological2015,PlatMagnetization2015,RanHidden2024,RanCubic2024,RanPhase2024}, and it could provide necessary theoretical and numerical guidance for the actively on-going experiments on frustrated quantum magnets and synthetic quantum simulators such as Rydberg atom arrays. The hard constraint of the QDM/QLM is naturally enforced by the Rydberg blockade mechanism, which prevents the simultaneous excitation of more than one atom within a specified radius. The mapping of the QDM and QLM on triangular~\cite{ZYan2022,Samajdar:2020hsw,zeng2025quantum} and kagome lattices~\cite{Semeghini21,Giudici2022Dynamical,Verresen2021Prediction,Verresen2022Unifying} to Rydberg atom arrays has facilitated the investigation of novel quantum phases, such as the $\mathbb{Z}_2$ spin liquid, in frustrated systems. The Rydberg atoms placed on the site of the kagome lattice can be mapped to the triangular dimer model, as we discussed in the present work. However, the kagome QDM, which corresponds to Rydberg atoms positioned on the links of the kagome lattice (ruby lattice)~\cite{Semeghini21,Giudici2022Dynamical,Verresen2021Prediction,Verresen2022Unifying}, remains to be explored. Previous ED study~\cite{Hwang2024Vison} are limited to small system sizes, potentially failing to capture quantum criticality and and transition. Our DMRG method offers a promising approach to overcoming these challenges and filling this research gap in future investigations.

The idea of symmetrizing constraint and the associated dynamic storage of quantum numbers can be readily applied on other constrained models, such as PXP model~\cite{Lin_Quantum_2020,Huang_Stability_2021}, which also describes synthetic quantum simulators. Our method can also serve as a useful tool to detect the topological double semion order in kagome dimer model with an extra term~\cite{oliver2014double}, where QMC would face the sign problem.

\begin{acknowledgments}
We thank Menghan Song, Hongyu Lu, and Min Long for valuable discussions on the DMRG implementation. We thank Fabien Alet, Sylvain Capponi, and Junchen Rong for the inspiring discussion on the physics of constrained lattice models, and we are in particular grateful for the insightful suggestions from Fabien Alet on the narrative and presentation of the manuscript. TTW, XXR and ZYM acknowledge the support from the Research Grants Council (RGC) of Hong Kong (Project Nos. 17301721, AoE/P-701/20, 17309822, HKU C7037-22GF, 17302223, 17301924), the ANR/RGC Joint Research Scheme sponsored by RGC of Hong Kong and French National Research Agency (Project No. A\_HKU703/22). We thank the HPC2021 system under the Information Technology Services, The University of Hong Kong, as well as the Beijing PARATERA Tech CO., Ltd. (URL: https://cloud.paratera.com) for providing HPC resources that have contributed to the research results reported in this paper.
\end{acknowledgments}

\appendix

\section{Implementation of local symmetries in MPO and MPS} \label{sec:DMRG_detail}

The total Hilbert space is defined as $\mathscr{H}=\bigotimes_b \mathscr{H}_b$, where each local Hilbert space on bond $b$ is a direct sum of empty and occupied dimer states, i.e., $\mathscr{H}_b=\ket{0}_b\oplus\ket{1}_b$. The conservation of the numbers of dimers can be viewed of as $N_\textrm{site}$ independent $\mathrm{U}(1)$ symmetries. Let $\vec{\theta}=(\theta_1,\cdots,\theta_{N_\textrm{site}})$ and $\hat{\vec{N}}=(\hat{n}_1,\cdots,\hat{n}_{N_\textrm{site}})$ be two $N_\textrm{site}$-dimensional vectors, where each $\theta_v$ is an angle from 0 to $2\pi$, and operator $\hat{n}_v=\sum_{b\in v}\hat{n}_b$ counts the number of dimers on bond connected to site $v$. The basis states of local Hilbert space $\mathscr{H}_b$ on bond $b$ transform under the $\mathrm{U}(1)$ rotations as,
$$e^{i\vec{\theta}\cdot\hat{\vec{N}}}\ket{0}_b=\ket{0}_b \text{, and }$$
$$e^{i\vec{\theta}\cdot\hat{\vec{N}}}\ket{1}_b=e^{i\vec{\theta}\cdot\vec{\mu}_b}\ket{1}_b,$$
where elements in vectors $\vec{\mu}_b=(0,\cdots,1,0,\cdots,1,0,\cdots,0)$ equal to 1 for the two sites $v$ connected to bond $b$ and 0 otherwise, i.e.,
$$[\vec{\mu}_b]_v=
\begin{cases}
    1,& \text{if } b\in v \\
    0,              & \text{otherwise}
\end{cases}
$$

For QDM, the constrained Hilbert space $\mathscr{H}_\textrm{QDM}\subset\mathscr{H}$ contains wave functions that are superpositions of product states under the one dimer per site constraint,
$$\ket{\Psi}=\sum_{\{\sigma_b=0,1\}}a_{\{\sigma_b\}}\bigotimes_b\ket{\sigma_b},$$
summing over states where $\sum_b\sigma_b\vec{\mu}_b=(1,1,\cdots,1)$.

The Hamiltonian commutes with the transformation $e^{i\vec{\theta}\cdot\hat{\vec{N}}}$. One has
$$\bra{\{\sigma_b\}}H\ket{\{\sigma'_b\}}\neq0 \text{, only if } \sum_b(\sigma_b-\sigma'_b)\vec{\mu}_b=(0,0,\cdots,0)$$

Therefore, we can track these quantum numbers as $N_\textrm{site}$-dimensional vector for MPO and MPS.

\begin{figure}[htp!]
	\centering
	\includegraphics[width=\columnwidth]{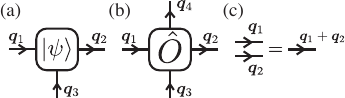}
	\caption{\textbf{Schematics for local tensors} Local tensors in (a) MPS and (b) MPO. The arrows indicate the flows of quantum numbers (vector) $\vec{q}$. The vertical (horizontal) arrows represent the physical (virtual) spaces. The quantum numbers add up respectively when the spaces are fused.}
	\label{fig:fig_tensor}
\end{figure}

Fig.~\ref{fig:fig_tensor} shows schematics for local tensors. Following the directions of the arrows in (a) and (b), one can mark each dimension with vector so that nonzero elements in each local tensor of an MPS (MPO) locate only in blocks where $\vec{q}_1+\vec{q}_3=\vec{q}_2 \, (+\vec{q}_4)$, where the quantum numbers at the physical dimensions is defined as $\sigma_b\vec{\mu}_b$.

\section{Ground state result}
\label{sec:Ground_state_result}
\begin{figure}[htp!]
	\centering
	\includegraphics[width=\columnwidth]{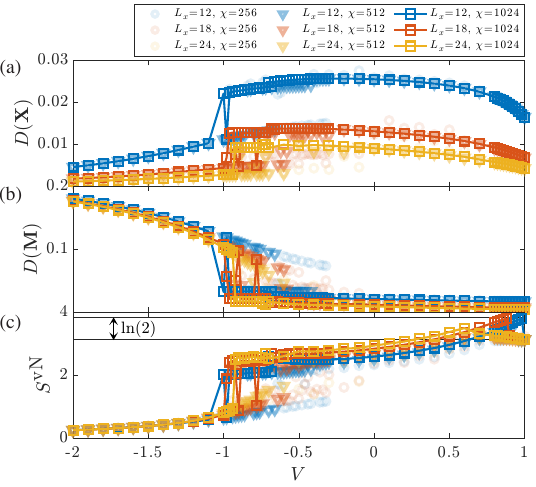}
	\caption{\textbf{Observables measured from MPS.} Result of dimmer correlation at (a) $\vec{X}$ and (b) $\vec{M}$ points, and (c) von-Neumann entanglement entropy. The three panels share the same legend on the top.}
	\label{fig:fig_GS}
\end{figure}

In this section, we present observables obtained from DMRG across the phase diagram.

The dimmer-dimmer correlations $D(\vec{X})$ and $D(\vec{M})$ are measured using the resulting MPS with the definition mentioned in the main text, except the $D(\vec{M})$ are computed for horizontal dimers, which is favored in the DMRG computation for negative $V$.

As shown in Fig.~\ref{fig:fig_GS} (a) \& (b), $D(\vec{X})$ ($D(\vec{M})$) is the peak in the Brillouin zone, serving as an order parameter for the VBS (columinar) phase. At around $V=-1$, there is a jump for both order parameters, which suggests a first-order transition between the two phases.

The entanglement entropy (EE) exhibits a sharp increase following the first-order transition and continues to rise with increasing $V$.

\begin{figure}[htp!]
	\centering
	\includegraphics[width=\columnwidth]{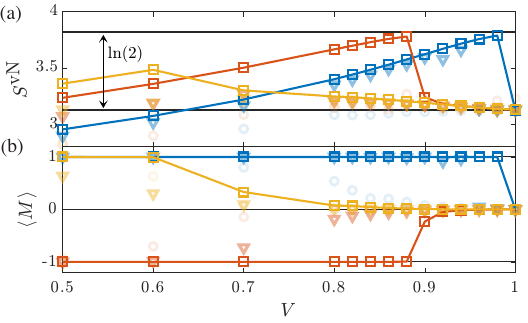}
	\caption{\textbf{Entanglement entropy and topological sector expectation value.} Result of (a) entanglement entropy and (b) the expectation value of operator $M$ defining the two topological sectors for $V$ from 0.5 to 1. The legend is shown on the top of Fig.~\ref{fig:fig_GS}.}
	\label{fig:fig_S_topo}
\end{figure}

Fig.~\ref{fig:fig_S_topo} presents a zoomed-in view of the EE and the expectation value of $M$ in the vicinity of the spin liquid phase. For small values of $V$, the DMRG algorithm correctly identifies the ground state within the appropriate topological sector of the VBS phase: specifically, $M = +1$ ($-1$) when $L_x/2$ is even (odd). Within the spin liquid phase, the two topological sectors become nearly degenerate. For states deep inside this phase, the DMRG algorithm tends to converge to a superposition of the two sectors, resulting in an average value of $\langle M \rangle$ that deviates from $\pm 1$ to around 0, that is the MES mentioned in the main text~\cite{Zhang_Quasiparticle_2012,Jiang_Identifying_2012,zhaoMeasuring2022}, accompanied by a characteristic reduction in EE.

\newpage

\bibliographystyle{longapsrev4-2}
\bibliography{bibtex}
\end{document}